\documentclass[11pt]{article}
\usepackage{amssymb, amsmath}
\begin{document}

\makeatletter
\@addtoreset{equation}{section}
\def\theequation{\thesection.\arabic{equation}}
\def\@maketitle{\newpage
 \null
 {\normalsize \tt \begin{flushright} 
  \begin{tabular}[t]{l} \@date  
  \end{tabular}
 \end{flushright}}
 \begin{center} 
 \vskip 2em
 {\LARGE \@title \par} \vskip 1.5em {\large \lineskip .5em \begin{tabular}[t]{c}\@author 
 \end{tabular}\par} 
 \end{center}
 \par
 \vskip 1.5em} 
\makeatother
\topmargin=-1cm
\oddsidemargin=1.5cm
\evensidemargin=-.0cm
\textwidth=15.5cm
\textheight=22cm
\setlength{\baselineskip}{16pt}
\title{
Study of the AdS$_2$/CFT$_1$ Correspondence with the Contribution from the Weyl Anomaly
}
\author{
Ryuichi~{\sc Nakayama}\thanks{nakayama@particle.sci.hokudai.ac.jp} \ \  \  and \ \ \ Tomotaka~{\sc Suzuki}\thanks{t-suzuki@particle.sci.hokudai.ac.jp} 
       \\[1cm]
{\small
    Division of Physics, Graduate School of Science,} \\
{\small
           Hokkaido University, Sapporo 060-0810, Japan}
}
\date{
EPHOU-17-001  \\
}
%
%
\maketitle

\begin{abstract} 
In this paper we will consider the Almheiri-Polchinski model of the AdS$_2$ back reaction coupled with Liouville field, which is  necessary for quantum consistency. 
 In this model, the Liouville field is determined classically by a bulk conformal transformation. 
The boundary time is also reparametrized by this transformation. 
It is shown that the  on-shell action on the boundary for the gravity sector is given by a bulk integral containing the Liouville field. This integral stems from  Weyl anomaly and is SL(2,R) invariant.   
A prescription is given for computing correlation functions of the operators dual to massless scalars.  
The generating function of the correlation functions of these operators 
is given by a sum of matter action and the bulk integral containing the Liouville field. The latter integral leads to extra  contributions to $n(\geq 6)$ point functions. 
\end{abstract}
\newpage
\setlength{\baselineskip}{18pt}

\newcommand{\bm}[1]{\mbox{\boldmath $#1$}}
\section{Introduction}
\hspace{5mm}
The Anti-de-Sitter/Conformal Field Theory (AdS/CFT) correspondence\cite{Maldacena}\cite{Witten}\cite{GKP} provides a relationship between effective gauge theories of the brane dynamics and string theory on the near-horizon AdS gravity. Two dimensional nearly Anti-de Sitter gravity (NAdS$_2$) appears in higher dimensional near extremal black hole geometry.\cite{ads21}\cite{ads22}\cite{ads23}\cite{ads24}\cite{Finite}\cite{Finite2}\cite{AP}\cite{C} Gravity in AdS$_2$ space is difficult to understand, because there are no finite energy excitations above the vacuum.\cite{Finite}   Asymptotic symmetry of the AdS$_2$ space is the reparametrization of the boundary time $\tau$. 
However, it was argued that it is spontaneously broken to SL(2,R) symmetry by AdS$_2$ space itself, and explicitly because we need to deform the boundary due to the back reaction of matter fields on the geometry. 
In \cite{AP} a toy model, Almheiri-Polchinski model (A-P model), based on Jackiw-Teitelboim (J-T) model was introduced to study back reaction to AdS$_2$. It was concluded that the time variable on the boundary becomes a dynamical variable and its dynamics is governed by a Schwarzian-derivative action. In \cite{Jensen}, \cite{MSY}, \cite{EMV}, \cite{AK} this model was further analysed based on the interest related to the SYK model.\cite{Kitaev}\cite{MS} In \cite{MSY} a prescription was proposed to integrate correlation functions of operators on the boundary over reparametrizations of the  boundary time by using a Schwarzian action as the effective action. 

The AdS$_{d+1}$/CFT$_d$ correspondence is expected to provide a precise definition of quantum gravity in terms of a  boundary non-gravitational field theory, and the quantum gravity must  include the effect of Weyl anomaly. Among quantum gravity theories the case of  two-dimensional one is well-studied. Although in two-dimensional gravity there are no physical degrees of freedom in the gravity sector, the conformal mode is related to the Weyl  anomaly and plays an important role.  
This anomaly in the bulk is represented by Polyakov action and in the conformal gauge it reduces to the local  Liouville action. 
Although this anomaly is a quantum effect,  this is already included in the definition of the measure for the path integral of the bulk gravity theory. In the study of holographic dictionary Liouville field is to be treated semiclasically.

In the literature  a boundary interaction and  quantum effects due to the Weyl anomaly are studied in \cite{EMV}, where the Polyakov action is not introduced from the beginning, but added only later. 
The purpose of this paper is to show that even if the Liouville field is present, the principle of holography still applies. 
In this paper we will start with a 2d model of gravity which is coupled to the Polyakov action in order to make the 2d gravity theory  quantum mechanically consistent, and investigate the effect of including Liouville action into the A-P model on the dynamical reparametrization of the boundary, and on the correlation functions of operators on the boundary. Higher dimensional quantum gravity such as 4-d one is also interesting, but it is difficult to deal with the modes related to the anomaly and this is beyond the scope of this paper. 

For reparametrization of the boundary we will adopt a prescription which is disctinct from that in \cite{AP}\cite{MSY}\cite{EMV}. 
Boundary conditions on matter fields at the time-like boundary will introduce to the boundary theory an explicit dependence on a boundary time. Then by back reaction of matters the spacetime itself will be deformed and become dependent on time.  This will induce a conformal transformation $\zeta=z+i\tau \rightarrow F(\zeta)$ in the bulk, and the Liouville field is determined by this function $F(\zeta)$. 
The boundary time is also reparametrize as $\tau \rightarrow -iF(i\tau)$. Other fields such as dilaton and matter fields are also determined in terms of $F(\zeta)$. By substituting these solutions to the equations of motion into the action, the boundary on-shell action is obtained. 
Liouville action in AdS$_2$ space, however,  breaks conformal symmetry and 
the dilaton field which corresponds to the reparametrized boundary time is not simply related by a conformal transformation of the original dilaton field corresponding to the undeformed boundary time.\footnote{This is because the background metric is not a flat one but an AdS one.}  
Nonetheless, it can be shown that the constraint equation of 2d gravity can be solved and  by substitution of it into the action we will derive an on-shell action on the boundary. It is shown that the Schwarzian-derivative action found in \cite{MSY}\cite{EMV}\cite{Kitaev} does not appear. Instead, a new term (\ref{BCaction2}) given by a 2d bulk integral appears, which comes from the Liouville self-interaction in the bulk. 
This term is an effect of Weyl anomaly. The appearance of a bulk integral might be puzzling. 
As is well-known, however, the on-shell action for an interacting scalar field in AdS space is given by a bulk integral which corresponds to Witten diagrams.\cite{Witten} After the integration over the radial coordinate, the on-shell action is given by an integral over the boundary. 
Similarly, in the case of the Liouville-dilaton gravity,  integration over the radial coordinate of the effective action yields a boundary on-shell action. 

We will then give a prescription to obtain correlation functions of boundary operators which takes into account the back reaction from the massless scalars. 
The calculation can be done in a similar way to the case \cite{AP} without Liouville field. 
One difference is that in addition to the on-shell matter action the above mentioned bulk integral  gives rise to contributions to the correlation functions, which also break conformal symmetry. This is an effect of  the Weyl anomaly. 
In this way, the holographic principle also applies to 2d gravity with Liouville field.

The on-shell boundary action can be also derived for a black hole spacetime. In this case in addition to the bulk integral of Liouville field, an additional term related to the solution of the dilaton field appear. For a static black hole solution this term becomes a term linear in the temperature $T$, and an entropy $S$ which is linear in $T$ is obtained. 
This agrees with the result in \cite{MSY}. 

AdS$_2$ appears in near horizon geometries of various extremal black hole solutions as a factored geometry. 
Hence one could UV-complete the present model by embedding it into higher-dimensional AdS spacetimes.
In this paper, however, we will not consider UV completion of AdS$_2$ by higher dimensional asymptotically AdS spacetime. 

This paper is organized as follows. In sec.2 J-T model and nearly AdS$_2$/CFT$_1$ based on it are briefly reviewed. In sec.3 Liouville action is coupled to J-T action and solutions to the classical equations of motion and constraints for the stress tensor are studied. In sec.4 an on-shell action is obtained. This on-shell action is a 2d bulk integral of the solution for the Liouville field. In sec.5 the boundary on-shell action for a black hole is studied.  In sec.6 it is shown that correlation functions of operators dual to the massless scalar fields can be computed by imposing a condition (\ref{tildetautau}) which determines a form of the function $F(\zeta)$. 
Hence holography also applies to Liouville theory. We conclude with a summary in sec.7.

\section{Nearly AdS$_2$/CFT}
\hspace{5mm}
We start with Jackiw-Teitelboim (JT) model for two-dimensional AdS gravity.
\begin{eqnarray}
S_{\text{JT}} &=& \frac{1}{16\pi G} \, \int d^2x \, \sqrt{g} \, \Phi \, ( R+2)+\frac{1}{8\pi G} \int d\tau \, \sqrt{\gamma} \,  \Phi \, K                                          \label{JT}
\end{eqnarray}
Here we work in Euclidean signature, and $g_{\mu\nu}$ is a bulk metric, $R$ the scalar curvature,\footnote{The sign convention of $R_{\mu\nu}$ is $R_{\mu\nu}=\partial_{\lambda}{\Gamma^{\lambda}}_{\mu\nu}+{\Gamma^{\lambda}}_{\mu\nu}{\Gamma^{\sigma}}_{\lambda\sigma}-\cdots$.}  and $\gamma$ an induced metric on the boundary at infinity. $K$ is an extrinsic curvature of the boundary. $G$ is a two-dimensional Newton constant, and $\Phi$ is a dilaton field. We set the AdS length $\ell_{AdS}=1$. 
The equation of motion for $\Phi$ is 
\begin{equation}
R=-2, 
\end{equation}
which can be solved as $g_{\mu\nu}=\hat{g}_{\mu\nu}$, where 
\begin{equation}
d\hat{s}_{\text{Poincar\'e}}^2= \hat{g}_{\mu\nu} \, dx^{\mu}dx^{\nu}= \frac{1}{z^2}(dz^2+d\tau^2) \label{refmetric}
\end{equation}
is the pure AdS$_2$ metric in the Poincar\'e patch. $z$ is the radial coordinate $(z>0)$, and  the boundary is located at $z=0$.

The equations of motion for $g_{\mu\nu}$ 
\begin{equation}
T_{\mu\nu}^{\Phi}= \frac{-1}{8\pi G}(\partial_{\mu} \Phi \, \partial_{\nu}\Phi-\hat{g}_{\mu\nu} (\hat{\nabla} \Phi)^2+g_{\mu\nu}\, \Phi)=0
\end{equation} 
are constraints and solved as 
\begin{equation}
\Phi= \frac{1}{z} \big(a+b\tau+c(\tau^2+z^2)\big).
\label{solPhi}
\end{equation}
Here $\hat{\nabla}_{\mu}$ is a covariant derivative associated with $\hat{g}_{\mu\nu}$, and $a$, $b$, $c$ are constants.  This fixes the boundary condition for $\Phi$.\footnote{For simplicity  we will set $b=c=0$.}
\begin{equation}
\Phi \rightarrow \frac{a}{z} \qquad (z \rightarrow 0) 
\label{BC}
\end{equation}

When a matter field $\chi$ is coupled to  (\ref{JT}), the dilaton field receives a back reaction from the matter 
\begin{equation}
T^{\Phi}_{\mu\nu}=-T^{\chi}_{\mu\nu},
\end{equation}
where $T^{\chi}_{\mu\nu}$ is a stress tensor for $\chi$, and the dilaton no longer satisfies the asymptotic behavior in pure AdS$_2$ (\ref{BC}). To reimpose the boundary condition, it is necessary to deform the boundary by a reparametrization of the time variable $\tau \rightarrow u$. For details see \cite{AP}\cite{MSY}\cite{EMV}. 

In the above prescription Weyl anomaly is not properly taken into account. 
In the next section we will incorporate a Liouville field into the system.

\section{Coupling to Liouville Field}
\hspace{5mm}
Two dimensional gravity has Weyl anomaly and to take account of this anomaly we need to introduce a dynamics of the conformal factor of the metric tensor. Let us introduce a conformal factor into the metric.
\begin{equation}
g_{\mu\nu}= e^{2\rho} \, \hat{g}_{\mu\nu},    \label{referencemetric}
\end{equation}
where $\hat{g}_{\mu\nu}$ is a reference metric (\ref{refmetric}) and $\rho$ is a Liouville field. Then the JT action (\ref{JT}) depends on $\rho$. 
When the massless matter fields are comformally coupled to gravity, the matter action does not depend on the conformal factor of the metric. However, if the matter fields are integrated out in the path integral, then the Polyakov action appears. 
\begin{equation}
S_{\text{P}}=\frac{-1}{4\pi  \alpha^2} \int d^2 x \sqrt{g} \, R \, \frac{1}{\square} \, R \label{Paction}
\end{equation}
Here $\alpha$ is a constant related to  the central charge of matter fields,  $C_M=N$ ($N$ is the number of scalar fields),  as
\begin{equation}
\alpha^2=\frac{3}{C_M-24}. \label{centralcharge}
\end{equation}
The appearance of (\ref{Paction})  is due to the fact that  the integration measure for the matter and other fields defined with respect to the metric $g_{\mu\nu}$ is not Weyl invariant.\cite{DK}\cite{FD} To rewrite (\ref{Paction}) into a local form it is necessary to split the metric tensor into a product of a reference metric and a conformal factor as in (\ref{JT}). 
For consistency the partition function of the gravity theory must be invariant under the following Weyl transformation, which is just a redundancy in (\ref{referencemetric}).
\begin{equation}
\hat{g}_{\mu\nu} \rightarrow e^{\sigma} \, \hat{g}_{\mu\nu}, \qquad \rho \rightarrow \rho-\frac{1}{2} \, \sigma \label{refWeyl}
\end{equation}
This requires the cancellation of the total conformal anomaly for $\hat{g}_{\mu\nu}$.
By substituting (\ref{referencemetric}) into (\ref{Paction}) 
 the Polyakov action is transformed into 
\begin{equation}
S_{\text{P}}(\rho,\hat{g})= -\frac{1}{8\pi \alpha^2} \int d^2 x \sqrt{\hat{g}} \, \Big((\hat{\nabla} \rho)^2+\hat{R} \rho+\mu e^{2\rho}\Big)-\frac{1}{4\pi \alpha^2} \int d\tau \, \sqrt{\hat{\gamma}} \, \rho \, \hat{K}. \label{PL}
\end{equation}
Here a boundary term is introduced. A cosmological term is also introduced and $\mu$ is a constant to be determined later by requiring consistency of the equations of motion. Its value can be adjusted by a local counterterm, because the anomaly is defined in terms of (\ref{Paction}), which is computed as a functional determinant. 

We will add this action to the Jackiw-Teitelboim action (\ref{JT}). By substituting (\ref{referencemetric}) into (\ref{JT}) and adding a suitable boundary term to make a variational problem well-defined, we obtain an action for Liouville-Dilaton gravity. 
\begin{eqnarray}
S_{LD} &=& \frac{1}{16\pi G} \, \int d^2x \sqrt{\hat{g}} \, \Big( \Phi(\hat{R}+2e^{2\rho})+2 \hat{\nabla}\Phi \cdot \hat{\nabla}\rho \Big)+\frac{1}{8\pi G} \, \int d\tau \, \sqrt{\hat{\gamma}} \, \Phi \, \hat{K} \nonumber \\
&&-\frac{1}{8\pi \alpha^2} \int d^2x \sqrt{\hat{g}} \, \Big( (\hat{\nabla}\rho)^2+\hat{R}\rho+\mu e^{2\rho}\Big)-\frac{1}{4\pi \alpha^2} \int d\tau \sqrt{\hat{\gamma}} \rho \hat{K}
\label{LD}
\end{eqnarray}
Furthermore matter action must be added. For real free massless scalar fields $\chi^i$ $(i=1, \dots N)$, it reads
\begin{equation}
S_M = \int d^2x \, \sqrt{\hat{g}} \, \sum_{i=1}^N  \frac{1}{2} (\hat{\nabla} \chi^i)^2.
\end{equation}
Since the action is rewritten in terms of a reference metric $\hat{g}_{\mu\nu}$, we may also transform the measure of the functional integral into a new one defined with respect to $\hat{g}_{\mu\nu}$. In this transformation a Jacobian will appear. It is assumed that the effect of this Jacobian is to simply modify the parameter $\alpha$ in (\ref{LD}). Actually, the modification of this value is not important here, as long as $C_M$ is large. \cite{DK}

The total action $S=S_{LD}+S_M$ by itself is in fact not invariant under the Weyl transformation (\ref{refWeyl}). 
However, the functional integral is invariant, 
provided a functional measure for fields in the path integral defined with respect to $\hat{g}$ transforms in the following way. 
\begin{equation}
[{\cal D}\rho]_{e^{\sigma} \, \hat{g}} 
[{\cal D}\chi^i]_{e^{\sigma} \, \hat{g}} 
[{\cal D}\text{(ghost)}]_{e^{\sigma} \, \hat{g}} 
=e^{S_P(\sigma/2, \, \hat{g})} 
\, [{\cal D}\rho]_{\hat{g}} 
[{\cal D}\chi^i]_{\hat{g}} 
[{\cal D}\text{(ghost)}]_{\hat{g}}  \label{changeMeasure}
\end{equation}
This prescription worked in $d<1$ gravity.\cite{DK} This is also assumed to be the case here. 
To recapitulate, the introduction of the Liouville field and functional integration over it  is necessary for consistent treatment of the Weyl anomaly. 

Let us now turn to the analysis of the equations of motion.
By variation of $\Phi$ and $\rho$ we obtain equations, 
\begin{eqnarray}
&&\hat{R}+2e^{2\rho}-2\hat{\nabla}^2 \, \rho=0, \\
&& \frac{1}{16\pi G} \, \Big(\hat{\nabla}^2 \, \Phi-2e^{2\rho} \, \Phi\Big)-\frac{1}{8\pi \alpha^2} \, \Big( \hat{\nabla}^2\rho-\frac{1}{2} \hat{R}-\mu \, e^{2\rho}\Big)=0.
\end{eqnarray}
The scalar curvature of (\ref{refmetric}) is $\hat{R}=-2$. 
If we set $\mu=1$, these equations are decomposed as 
\begin{eqnarray}
&&\hat{\nabla}^2 \, \rho-e^{2\rho}+1=0, \label{rhoeq}\\
&& \hat{\nabla}^2 \, \Phi-2e^{2\rho} \, \Phi=0. \label{Phieq}
\end{eqnarray}
Henceforth we will choose this value for $\mu$. 
Matter equations of motion are given by 
\begin{equation}
\hat{\nabla}^2 \, \chi^i=0. \label{chieq}
\end{equation}

The constraint equations which comes from $\hat{g}_{\mu\nu}$ are 
\begin{eqnarray}
\frac{1}{8\pi G}\Big(  \hat{\nabla}_{\mu} \, \hat{\nabla}_{\nu}\Phi-\hat{\nabla}_{\mu} \Phi \,  \hat{\nabla}_{\nu} \rho 
-\hat{\nabla}_{\nu} \Phi \,  \hat{\nabla}_{\mu} \rho 
-\text{trace}\Big)=T^{\rho}_{\mu\nu}+T^{\chi}_{\mu\nu}
\label{constraint}
\end{eqnarray}
in the case of the variation of the traceless part of the metric. 
Here $T^{\rho}$ and $T^{\chi}$ are traceless energy momentum tensors for $\rho$ and $\chi$.
\begin{eqnarray}
T^{\rho}_{\mu\nu} &=& \frac{-1}{4\pi \alpha^2} \, \Big(\partial_{\mu}\rho \partial_{\nu} \rho-\hat{\nabla}_{\mu}\hat{\nabla}_{\nu}\rho-\text{trace}\Big), \\
T^{\chi}_{\mu\nu} &=& \sum_{i=1}^N\partial_{\mu}\chi^i \partial_{\nu} \chi^i-\text{trace}
\end{eqnarray}
For the trace part, we have
\begin{eqnarray}
\frac{1}{2}\hat{\nabla}^2\Phi-\Phi-\Phi \hat{\nabla}^2\rho+\frac{1}{\alpha^2} \, G \, \Big[e^{2\rho}-\hat{\nabla}^2\rho\Big]=0.  \label{traceeq}
\end{eqnarray}
This last equation is actually  not complete, because the variation of the functional measure for $\rho$, (\ref{changeMeasure}), is not taken into account. From (\ref{changeMeasure}) we obtain an extra term $G/\alpha^2$ on the right-hand side of (\ref{traceeq}). Then  (\ref{traceeq}) modified in this way is consistent with (\ref{rhoeq}), (\ref{Phieq}) and (\ref{chieq}). 
Hence with the choice $\mu=1$, independent equations of motion are (\ref{rhoeq}), (\ref{Phieq}), (\ref{chieq}) and (\ref{constraint}).

Now, we will set $\chi^i=0$ for simplicity. If we choose a solution $\rho=0$ for (\ref{rhoeq}), then the solution to (\ref{Phieq}) is given by (\ref{solPhi}). On the other hand, a general solution to (\ref{rhoeq}) which does not change the boundary at $z=0$ is 
\begin{equation}
\rho_F(z,\tau)\equiv \frac{1}{2}\, \log \frac{4z^2 F'(i\tau+z)F'(i\tau-z)}{[ F(i\tau+z)-F(i\tau-z)]^2}.  \label{rhoF}
\end{equation}
Here $F(x)$ is a real and monotonic  function.\footnote{ In Euclidean space $F(x)$ must be also an odd function in order for $\rho_F$ to be real. Then out of SL(2,R) transformations only a dilatation satisfies this condition. In Lorentzian space with $t=-i \tau$ this condition is not necessary. Henceforth this condition is not imposed on $F(x)$, even if $\rho$ becomes complex, because this problem disappears after Wick rotation.} A map $z+i\tau \rightarrow F(z+i\tau)$ is a conformal transformation in the bulk.  This is different from the prescription of deforming the regularized boundary at $(z=\varepsilon,\tau)$ to $(z=\varepsilon \, F'(i\tau),-iF(i\tau))$, while keeping the bulk metric. \cite{AP}\cite{MSY}\cite{EMV}  In this paper, the position of the boundary is fixed at $z=\varepsilon$ with a small $\varepsilon$. 

On the other hand, the solution $\Phi$ to (\ref{Phieq}) and (\ref{constraint})  is not obtained by a conformal transformation $i\tau  \pm z \rightarrow F(i\tau \pm z)$ of (\ref{solPhi}). 
Actually, if we define 
\begin{multline}
\hat{\Phi} (z,\tau)\equiv \frac{1}{F(i\tau+z)-F(i\tau-z)} \, \Big[2a-ib \{F(i\tau+z)+F(i\tau-z) \}\\ 
-(c/2) (F(i\tau+z)+F(i\tau-z))^2+ (c/2)(F(i\tau+z)-F(i\tau-z))^2 \Big], \label{PhiF}
\end{multline}
it turns out the constraint (\ref{constraint}) is not satisfied. This is because the background metric (\ref{refmetric}) is not a flat metric.  Indeed, if we introduce complex coordinates $\zeta$ and $\bar{\zeta}$ by
\begin{equation}
\zeta=z+i\tau, \quad \overline{\zeta}=z-i\tau, \label{zetazetabar}
\end{equation}
we obtain
\begin{eqnarray}
\frac{-1}{8\pi G}\Big(  \hat{\nabla}^2_{\zeta} \, \hat{\Phi}-2\, \hat{\nabla}_{\zeta} \hat{\Phi} \,  \hat{\nabla}_{\zeta} \, \rho_F
\Big)+T^{\rho}_{\zeta\zeta}|_{\rho=\rho_F}=\frac{1}{8\pi \alpha^2} \, \{F(\zeta), \zeta\}.
\label{Tzz}
\end{eqnarray}
Here 
\begin{equation}
S(\zeta) \equiv \{F(\zeta),\zeta \} \equiv \frac{F'''(\zeta)}{F'(\zeta)}-\frac{3}{2}\, \Big(\frac{F''(\zeta)}{F'(\zeta)}\Big)^2
\end{equation}
 is a Schwarzian derivative. Hence only for SL(2,R) transformation $\zeta \rightarrow F(\zeta)=(a_1 \zeta+a_2)/(a_3 \zeta +a_4)$, $(a_1a_4-a_2a_3=1)$, (\ref{rhoF}) and (\ref{PhiF}) satisfiy the constraint (\ref{constraint}). 

Instead, given (\ref{rhoF})  we can solve  (\ref{Phieq}) and constraint (\ref{constraint}).
We will now solve the following equations.
\begin{eqnarray}
\hat{\nabla}_{\zeta}^2 \Phi -2 \hat{\nabla}_{\zeta}\, \rho_F \cdot \hat{\nabla}_{\zeta} \Phi&=&-\frac{2G}{\alpha^2} \, \Big((\partial_{\zeta}\rho_F)^2-\hat{\nabla}_{\zeta}^2\rho_F \Big), \label{Phirho}\\
  (\hat{\nabla}^2-2e^{2\rho_F})\, \Phi &=& 0
\end{eqnarray}
The right-hand side of (\ref{Phirho}) is given by $G\alpha^{-2}S(\zeta)$.
By substituting $\rho_F$, the above equations are made more explicit.
\begin{eqnarray}
\partial_{\zeta}^2\Phi+\Big( \frac{1}{z}-\frac{2}{3}zS(i\tau)-\frac{1}{3} z^2S'(i\tau)+O(z^3)\Big) \partial_{\zeta}\Phi= \frac{G}{\alpha^2}S(\zeta), \label{contR}\\
\Big( z^2(\partial_z^2+\partial_{\tau}^2)-2\Big) \, \Phi =\Big(\frac{4}{3}z^2S(i\tau)+O(z^2)\Big) \, \Phi \label{eomR}
\end{eqnarray}
By defining 
\begin{equation}
\Phi(z,\tau)=\sum_{n=-1}^{\infty} z^n \, \Phi_n(\tau),  \label{Phiexp}
\end{equation}
and solving the above equations, we have $\Phi_0(\tau)=0$ and a set of equations which determine $\Phi_{-1}(\tau)$ and $\Phi_1(\tau)$:
\begin{eqnarray}
&& \frac{d^3\Phi_{-1}}{d\tau^3} \, -2S(i\tau)\frac{d\Phi_{-1}}{d\tau}-\frac{4}{3}\frac{dS(i\tau)}{d\tau}\Phi_{-1}=0, \label{Phim1} \\
&& \Phi_1=\frac{1}{2}\frac{d^2\Phi_{-1}}{d\tau^2}-\frac{2}{3}S(i\tau) \Phi_{-1}.  \label{Phi1}
\end{eqnarray}
Explicit solutions to these equations are not obtained yet. Three independent solutions which correspond to $F(\zeta)=\zeta+\epsilon(\zeta)$, which is infinitesimally close to an  identity conformal transformation, are the following. 
\begin{eqnarray}
\Phi^{(0)}_{-1}(\tau) &=& 1+\frac{4}{3} \,  \epsilon'(i\tau)+\dots, \nonumber \\
\Phi^{(1)}_{-1}(\tau) &=& \tau+2i\epsilon(i\tau)+\frac{4}{3} \, \tau \epsilon'(i\tau)+\dots, \nonumber \\
\Phi^{(2)}_{-1}(\tau) &=& \tau^2+\frac{4}{3} \, \tau^2\epsilon'(i\tau)+4i\tau\epsilon(i\tau)-4i\int^{\tau} \epsilon(i\tau_1)d\tau_1+\dots
\end{eqnarray}
Here $\epsilon'(\zeta)=(d/d\zeta) \epsilon(\zeta)$. Hence the solution $\Phi$ breaks SL(2,R).

\section{On-shell Action for Gravity Sector}
\hspace{5mm}
On-shell action is obtained by substituting solutions (\ref{rhoF}) and (\ref{PhiF}) into the action $S_{LD}$ (\ref{LD}). 
After using equations of motion (\ref{rhoeq}) it is simplified as follows. 
\begin{multline}
S^{\text{onshell}}_{LD}
= \frac{-1}{8\pi G}\int d\tau \,   \Phi \, \partial_{z} \rho_F|_{z=\varepsilon}+\frac{1}{8\pi G} \int d\tau \,  \frac{1}{\varepsilon} \,  \Phi \, \hat{K}|_{z=\varepsilon} \\
-\frac{1}{4\pi \alpha^2}\int d\tau  \,\frac{1}{\varepsilon} \,  \rho_F \, \hat{K}|_{z=\varepsilon}
 +\int d\tau \, \frac{1}{8\pi \alpha^2} \, \rho_F \, \partial_z \rho_F|_{z=\varepsilon}\\
-\frac{1}{8\pi \alpha^2} \int d\tau  \,  \int_{\varepsilon }^{\infty} \frac{dz}{z^2} \, \Big(-\rho_F +(1-\rho_F) \, e^{2\rho_F}\Big)
 \label{on-shell}
\end{multline}
Here the location of the boundary is regularized as $z=\varepsilon$, 
where $\varepsilon$ is a small positive constant. 
Notice that the boundary is flat. The extrinsic curvature at the boundary is $\hat{K}=1$. 
Since near the boundary $z \rightarrow 0$, $\rho_F$ and $\Phi$ behave as 
\begin{eqnarray}
\rho_F(z,\tau) &\sim& \frac{1}{3}z^2 \, \{F(i\tau), i\tau \}, \\
\Phi(z,\tau)  &\sim& \frac{1}{z} \, \Phi_{-1}(\tau) +z \, \Phi_1(\tau), \label{PhiFA}
\end{eqnarray} 
the third and fourth terms in (\ref{on-shell}) vanish and the first integral is finite as $\varepsilon \rightarrow 0$. The last bulk integral diverges in the UV limit $\varepsilon \rightarrow 0$. However,  the  divergence is proportional to $\int d\tau\int_{\varepsilon}^{\infty} dz \sqrt{\hat{g}}$ and behaves as a constant $\frac{1}{8\pi \alpha^2 \varepsilon} \int d\tau 1$ for a small, but finite $\varepsilon$.  
On the other hand, the second boundary term 
is divergent as $\varepsilon \rightarrow 0$ for the solution (\ref{PhiFA}).\footnote{In \cite{MSY}, this problem is handled by introducing a boundary-value problem  with condition $\Phi |_{\text{bndry}} = \frac{1}{\epsilon} \, \Phi_r(\tau)$ and making $\Phi_r$ independent of $F$. Then the divergence is a constant. From the equation of motion for $F$ in the Schwazian-derivative theory, the solution for $\Phi$ (\ref{solPhi}) is reproduced.}
\begin{equation}
\frac{1}{8\pi G}\int d\tau \Big[\frac{\Phi_{-1}(\tau)}{\varepsilon^2} 
+\Phi_1(\tau)\Big]
\end{equation}
The first term is large for small $\varepsilon$ and in general $\Phi_{-1}$ is not a constant. In a cut-off theory we cannot renormalize this divergence. We can, however, make such a term a constant one by imposing a condition
\begin{equation}
\Phi_{-1}(\tau) =a, 
\label{tildetautau}
\end{equation}
where $a$ is a constant.  
As will be discussed at the end of this section this condition limplies a trivial conformal transformation $F(\zeta)=\zeta$. 
In sec. 6 this condition will be imposed after turning on the source terms $j$  for the boundary operators dual to massless scalar fields.  Then $F(\zeta)$ will be a non-trivial function and depend on the source functions. 
 
Continuing the analysis of the on-shell action, with the condition (\ref{tildetautau}) we obtain the following on-shell action 
\begin{multline}
S^{\text{onshell}}_{LD}
= \frac{1}{12\pi G} \, \int_{-\infty}^{\infty} d\tau \, \Phi_{-1}(\tau) \{F(i\tau),i\tau\}+\frac{1}{8\pi G}\int_{-\infty}^{\infty}d\tau \,  \Phi_1(\tau)  \\
-\frac{1}{8\pi \alpha^2} \int_{-\infty}^{\infty} d\tau  \int_{0}^{\infty}dz \Big[\sqrt{\hat{g}} \, \Big(-\rho_F +(1-\rho_F) \, e^{2\rho_F}-1\Big)\Big]   +\text{const} \label{BCaction}
\end{multline}
The first term in (\ref{BCaction}) is the Schwarzian-derivative action \cite{Kitaev}\cite{MSY}\cite{EMV} with the extra factor $\Phi_{-1}$.  
There are two other terms in (\ref{BCaction}). The bulk integral stems from  the conformal anomaly.  Interestingly, it turns out that the first and second terms cancel out due to the relation (\ref{Phi1}). 
\begin{eqnarray}
S^{\text{onshell}}_{LD}
=-\frac{1}{8\pi \alpha^2} \int_{-\infty}^{\infty} d\tau  \int_{0}^{\infty}dz \Big[\sqrt{\hat{g}} \, \Big(-\rho_F +(1-\rho_F) \, e^{2\rho_F}-1\Big)\Big]   +\text{const} \label{BCaction2}
\end{eqnarray}
Hence the terms which break SL(2,R) symmetry canceled out and action (\ref{BCaction2})  is  invariant.\footnote{We note that (\ref{rhoF}) is invariant under SL(2,R).} 
This action appeared due to the Weyl anomaly, because there is a pre-factor $1/\alpha^2$.
The appearance of a bulk integral might be puzzling. 
As is well-known, however, the on-shell action for an interacting scalar field in AdS space is given by a bulk integral which corresponds to Witten diagrams.\cite{Witten} After the integration over the radial coordinate, the on-shell action is given by an integral over the boundary. 
Similarly, in the case of the Liouville-dilaton gravity,  integration on the radial coordinate of the effective action yields a boundary on-shell action.

Let us discuss  solutions to (\ref{tildetautau}), $\Phi_{-1}(\tau)=a$.  When  the matters $\chi^i=0$, by (\ref{Phim1}) the Schwarzian derivative $S(i\tau)$ must be a constant. 
Then up to SL(2,R) transformations the function $F(\zeta)$ is one of the following.
\begin{eqnarray}
F(\zeta) =\zeta, \quad
 \tanh (A\zeta), \quad
 \tan (A\zeta),
\end{eqnarray}
where $A$ is a real constant. 
For the first solution the on-shell action vanishes. 
The second solution corresponds to a black hole metric and the third for a special value of $A$ to an AdS$_2$ metric  in the  global coordinates. In these cases although the on-shell action does not vanish, it diverges inside the bulk. It is not known how to avoid this divergence, and  at present it is not clear if these solutions are meaningful. In the next section the black hole solution will be studied by choosing the black hole metric as the reference metric $\hat{g}$. 
In sec.6 this on-shell action is used for obtaining the generating function of correlation functions after turning on the matter field. 

 It will be interesting, if it is possible to evaluate the $z$-integral  in (\ref{BCaction2}) explicitly without carrying out  an expansion into a series.  This is not attempted in this paper. 
Instead here, this integral is studied for $F(\zeta)$ near the identity transformation. The result below  will be used in sec. 6. 
For the pure gravity in AdS$_2$ ($F(\zeta)=\zeta$), the Liouville field (\ref{rhoF}) is $\rho_F=0$. In this case action (\ref{BCaction2}) vanishes.  
For a perturbation $F(\zeta)=\zeta+\epsilon(\zeta)$, where $\epsilon(\zeta)$ is infinitesimal, (\ref{rhoF}) is expanded as 
\begin{equation}
\rho_F = \frac{1}{2}[ \epsilon'(\zeta)+\epsilon'(-\bar{\zeta})]-
\frac{1}{2z} \, [\epsilon(\zeta)-\epsilon(-\bar{\zeta})]
+ O(\epsilon^2).
\end{equation}
This vanishes as $z \rightarrow 0$. 
This also vanishes for $\epsilon(\zeta)=1,\zeta,\zeta^2$. 
Because the integrand of (\ref{BCaction})  behaves for small $\rho_F$ as
\begin{equation}
-1-\rho_F+(1-\rho_F) \, e^{2\rho_F} \sim 
-\frac{2}{3} \, (\rho_F)^3, \label{LDexp}
\end{equation}
the on-shell action (\ref{BCaction2}) is $O(\epsilon^3)$. 

\section{ AdS$_2$ Black Hole}
\hspace{5mm}
In sec.4 we obtained the on-shell action for the dynamical reparametrization. 
In this action the Schwarzian derivative term found in \cite {MSY}, \cite{EMV} does not appear, and furthermore a bulk 2d integral appears. 
If the on-shell action were also the same for  the AdS black hole, then the entropy of the black hole would vanish.

In this section we will consider an on-shell action in the case of an AdS$_2$ black hole. We will choose  the reference metric as $\hat{g}_{\mu\nu}$
\begin{equation}
d\hat{s}_{\text{BH}}^2 = \hat{g}_{\mu\nu} dx^{\mu}dx^{\nu}
= \frac{4\pi^2}{\beta^2 \, \sinh^2 \frac{2\pi z}{\beta}} \Big[ dz^2+d\tau^2\Big]. \label{BH}
\end{equation}
Here $\beta=\frac{1}{T}$
is an inverse temperature of the black hole. 
The solution to the Liouville equation (\ref{rhoeq}) in this case is given by 
\begin{equation}
\rho_F(z,\tau) = \frac{1}{2} \, \log \frac{\beta^2\sinh^2(2\pi z/\beta)  F'(\zeta)F'(-\bar{\zeta})}{\pi^2 [F(\zeta)-F(-\bar{\zeta})]^2}
\end{equation}
If $F(\zeta) \neq \tanh (\pi \zeta/\beta)$, then the spacetime is not static and fields are not in general periodic under $\tau \rightarrow \tau+\beta$. In this case Lorentzian time $t=-i\tau$ must be used and the integration region is $-\infty <t< \infty$. 
The equation of motion for $\Phi$ (\ref{Phieq}) and constraint (\ref{constraint}) are solved in a similar way to that in sec.3. 
It is, however, necessary to notice that  the stress tensor of Liouville field is modified to 
\begin{equation}
T^{\rho}_{\zeta\zeta}=\frac{1}{8\pi \alpha^2} \, \Big[\{F(\zeta),\zeta\}+\frac{2\pi^2}{\beta^2} \Big].
\end{equation}
By expanding $\Phi$ as (\ref{Phiexp}) we obtain equations (\ref{contR}) and (\ref{eomR}) with the right-hand side of (\ref{eomR}) modified due to the above change. The equations for $\Phi_{-1}$, $\Phi_0$ and $\Phi_1$ are the same as (\ref{Phim1}), (\ref{Phi1}) and $\Phi_0=0$.
The change in $T^{\rho}_{\zeta\zeta}$ only modifies $\Phi_2$ and higher coefficient functions.

The on-shell action is obtained by repeating the arguments in sec.4. 
A difference comes from the extrinsic curvature. For the metric (\ref{BH}) it is given by
\begin{equation}
\hat{K}= \cosh \frac{2\pi z}{\beta}=1+\frac{2\pi^2}{\beta^2} \, z^2+\cdots
\end{equation}
The second term in the near-boundary expansion above modifies the calculation of the second term of (\ref{on-shell}) 
\begin{eqnarray}
\frac{1}{8\pi G} \int_{-\infty}^{\infty} d\tau \frac{1}{\epsilon} \Phi \hat{K}|_{z=\epsilon}&=&\frac{1}{8\pi G}\int_{-\infty}^{\infty} d \tau \,  \frac{1}{\epsilon} \, (\frac{\Phi_{-1}}{\epsilon}+\Phi_1\epsilon+\dots)(1+\frac{2\pi^2}{\beta^2}\epsilon^2+\dots) \nonumber \\
&=& \frac{1}{8\pi G}\int_{-\infty}^{\infty} d \tau \, \Big(\frac{1}{\epsilon^2}\Phi_{-1}+\Phi_1+\frac{2\pi^2}{\beta^2}\Phi_{-1}+\dots \Big)
\end{eqnarray}
and a new term remains after the cancellation of $\Phi_0$ and $\Phi_1 \{F,i\tau\}$ in the first and second terms of (\ref{on-shell}). 
To summarize the on-shell action for a black hole is given by 
\begin{eqnarray}
S^{\text{onshell}}_{LD}
=\frac{\pi}{4G\beta^2}\int_{-\infty}^{\infty} dt \Phi_{-1}-\frac{1}{8\pi \alpha^2} \int_{-\infty}^{\infty} dt  \int_{0}^{\infty}dz \Big[\sqrt{\hat{g}} \, \Big(-\rho_F +(1-\rho_F) \, e^{2\rho_F}-1\Big)\Big]    \label{BCaction3}
\end{eqnarray}

For a black hole solution we have $F(\zeta)=\tanh \frac{\pi \zeta}{\beta}$ and the Euclidean time can be used. The Liouville field is trivial, $\rho_F(z,\tau)=0$ and the bulk integral vanishes. As for the first term by noting 
\begin{equation}
\{F(t),i \tau \}=-2 \, \Big(\frac{\pi}{\beta}\Big)^2 
\end{equation}
and using (\ref{Phim1}), we have 
\begin{equation}
\Phi_{-1}'''+\frac{4\pi^2}{\beta^2}\Phi_{-1}'=0.
\end{equation}
Then we can choose a solution $\Phi_{-1}=a $ (constant)
and the on-shell action is given by
\begin{equation}
S_{LD}^{\text{onshell}} = \frac{\pi \, a}{4G \, \beta}  =  \frac{\pi \, a}{4G} \, T.
\end{equation}
Then an entropy $S=\frac{a\pi}{2G} \, T$ is obtained. This linear-in-$T$ behavior coincides with the result of  \cite{MSY}.

When the spacetime is deformed away from the stationary black hole, the second term in (\ref{BCaction3}) is non-vanishing. In this case this term may be evaluated by expanding $F(z+t)$ in a power series of $z$. 
After integration over $z$, the on-shell action can be expanded in powers of the Schwarzian derivative and its derivatives. 
Up to now, however,  a systematic method for the expansion are not available.

\section{Correlation Functions of Scalar Operators}
\hspace{5mm}
When matter fields are turned on, it is  possible to regard a sum of (\ref{BCaction2}) and the matter action as the generating functional for correlation functions dual to the matter. 
When massless fields $\chi^i$ are turned on, the equation of motion for $\Phi$ (\ref{Phieq}) and the constraint equation (\ref{constraint})  are modified and so are their solutions.  We  show that  the condition  (\ref{tildetautau})
determines $F$, and captures the back reaction of the matter. This solution $F$ depends on the boundary sources $j$ of operators. Especially, when the boundary conditions $j$ for matter fields vanish, the solution $F$ coincides with the identity $F(\zeta)=\zeta$. 
By substituting the solution for non-vanishing $j$ into the on-shell action, the generating function for the correlation functions is obtained. In this way in the presence of Liouville field the principle of holography also applies and the partition function of 2d AdS$_2$ gravity in the semi-classical approximation yields a generating function of conformal quantum mechanics. 

In this section we will show that the constraint equation (\ref{constraint}) can be solved even in the presence of the matter fields and discuss computation of the correlation function of operator dual to a massless field. 
We can use (\ref{tildetautau}) to determine  the function $F(\zeta)$ and substitute this into the on-shell action. 
For simplicity we will keep only one massless field $\chi$ and set the boundary conditions of the other massless scalars to zero.

The solution to the Klein-Gordon equation $\hat{\nabla}^2\chi=0$ with boundary condition $\chi(z,\tau) \rightarrow j(\tau)$ as $z \rightarrow 0$ is given by
\begin{equation}
\chi(z,\tau) =\frac{1}{\pi} \, \int_{-\infty}^{\infty}\frac{z}{z^2+(\tau-\tau')^2} \, j(\tau')d\tau'
\end{equation}
This solution has an expansion
\begin{equation}
\chi(z,\tau)= \frac{1}{2\pi} \, \sum_{n=0}^{\infty} z^n \, j_n(\tau)
\end{equation}
with 
\begin{eqnarray}
j_0 (\tau) = 2\pi j(\tau),    \qquad 
j_1(\tau) = 2\int d\tau' \frac{P}{(\tau-\tau')^2}\, j(\tau')
\end{eqnarray}
Here $P$ represents for a principal value. 
The on-shell value of the matter action is
\begin{eqnarray}
S_M &=& -\int d\tau \frac{1}{2}\sqrt{\hat{g}} g^{zz} \chi\partial_z\chi|_{z=0} \nonumber \\
&=& -\frac{1}{2\pi} \, \int d\tau d\tau' j(\tau)\frac{P}{(\tau-\tau')^2} j(\tau') \label{SM1}
\end{eqnarray} 
In the new coordinates $(\tilde{z},\tilde{\tau})=((F(\zeta)-F(-\bar{\zeta}))/2, (F(\zeta)+F(-\bar{\zeta}))/2i)$, the source $j$ transforms as a scalar field 
$\tilde{j}(-iF(i\tau))=j(\tau)$, and the on-shell action is given by 
\begin{equation}
\tilde{S}_M= -\frac{1}{2\pi}\int d\tau_1 d\tau_2 F'(i\tau_1)F'(i\tau_2) \, \frac{P}{[-iF(i\tau_1)+iF(i\tau_2)]^2} j(\tau_1)j(\tau_2).
\end{equation}

When the solution to the Liouville equation is  choosen to be $\rho=\rho_F$ (\ref{rhoF}) instead of $\rho=0$,  the massless scalar is given by 
\begin{equation}
\tilde{\chi}(z,\tau) = \chi(\tilde{z},\tilde{\tau})=\frac{1}{2\pi}\sum_{n=0}^{\infty}\tilde{z}^n \, j_n(\tilde{\tau})=\frac{1}{2\pi}\sum_{n=0}^{\infty}z^n \, \tilde{j}_n(\tau).
\end{equation}
For example, 
\begin{equation}
\tilde{j}_0(\tau)=j_0(-iF(i\tau)), \qquad \tilde{j}_1(\tau) =F'(i\tau) \, j_1(-iF(i\tau)).
\end{equation}

The constraint equation for $\Phi$ (\ref{constraint}) and equation of motion for $\Phi$ (\ref{Phieq}) are given by 
\begin{eqnarray}
&& \hat{\nabla}_{\zeta}^2 \Phi -2 \hat{\nabla}_{\zeta} \Phi = 8\pi G \, T^{\rho}_{\zeta\zeta}|_{\rho=\rho_F}+8\pi G \,  (\partial_{\zeta}\tilde{\chi})^2, \\
&& \nabla^2\Phi - 2 \, e^{2\rho_F} \, \Phi=0.
\end{eqnarray}
The solution $\Phi$ is obtained as a series $\Phi=\sum_{n=-1}^{\infty} z^n\Phi_n(\tau)$. The first few coefficients are determined by the following equations. 
\begin{eqnarray}
&&\Phi_0 = 0, \\
&&\Phi''_{-1}-\frac{4}{3}S(i\tau)\Phi_{-1}=2\Phi_1, \label{1} \\
&& \Phi'_1=\frac{1}{3}S(i\tau) \, \Phi'_{-1}+\frac{G}{\pi} \tilde{j}_1\tilde{j}'_0. \label{2}
\end{eqnarray}
Here a prime stands for a derivative with respect to $\tau$. 
By combining (\ref{1}) and (\ref{2}) $\Phi_{-1}$ is obtained by solving
\begin{equation}
(\Phi_{-1})'''-2S(i\tau)(\Phi_{-1})'-\frac{4}{3}\frac{dS(i\tau) }{d\tau}\, \Phi_{-1}=\frac{2G}{\pi}\tilde{j}_1 \, \tilde{j}_0'.
\end{equation}
Then (\ref{1}) determines $\Phi_1$. 

By the procedure explained at the beginning of this section a generating functional for correlation functions of operators can be obtained.  
Equation (\ref{tildetautau}) defines a new time $\tilde{\tau}$. By (\ref{tildetautau})  $\Phi_{-1}=a$ and $F$ is determined. Up to $O(\tilde{j}^2)$ $F$ is given by 
\begin{multline}
F(i\tau)= i\tau+\frac{G}{2 a}\int d\tau_1 \int d \tau_2 \ j'(\tau_1)j'(\tau_2) \frac{P}{\tau_1-\tau_2} [(\tau-\tau_1)^3\theta(\tau-\tau_1)-(\tau-\tau_2)^3\theta(\tau-\tau_2)] \\
+O(j^4) \label{Ftau}
\end{multline}
This agrees with the result of \cite{AP} for the back reaction. 

The full boundary action is given after using (\ref{1}) by
\begin{eqnarray}
S_{LD}^{\text{onshell}}& =& -\frac{1}{8\pi \alpha^2} \int d\tau\int dz \sqrt{\hat{g}} \, [
-\rho_F+(1-\rho_F)e^{2\rho_F}-1]  +\tilde{S}_M  \label{bndry}
\end{eqnarray}
The two terms on the right-hand side are functionals of $F$. 
By substituting (\ref{Ftau}),  the generating functions of the operator dual to $\chi$ is obtained. In addition to a generating function from $\tilde{S}_M$, which was  obtained in \cite{AP}, a new contribution is obtained by inserting (\ref {rhoF}) into the bulk integral. Due to (\ref{LDexp}) this is a generating function of  $n(\geq 6$) point functions. 
This represents a contribution of the Weyl anomaly.  If $j(\tau)$ has a finite support, (\ref{bndry}) will be finite. To carry out the radial integral it is necessary to Wick rotate to Lorentzian time and then 
split the integration region into many pieces. The result is messy and will not be presented here.

\section{Summary}
\hspace{5mm}
In this paper Liouville field is introduced into AdS$_2$ gravity to take into account the effect of Weyl anomaly, 
and its effect on the nearly AdS$_2$/CFT$_1$ correspondence is analyzed. 
An on-shell boundary action is studied by changing not only the bulk time but also the bulk metric and by keeping the location of the boundary at $z=\varepsilon$. It is given by a bulk integral of some function of the Liouville field and is entirely due to the Weyl anomaly.  
This on-shell action is shown that even if Liouville filed is introduced, the ordinary prescription of AdS/CFT works and  
 correlation functions of boundary operators dual to matters can be  obtained.  The correlation functions receive contributions from the Weyl anomaly. 
It is interesting that although the Schwarzian-derivative action is absent from the boundary on-shell action, correlation functions similar to those in the theory without the Liouville field is obtained.  In this paper correlation functions of only  massless scalar fields are considered. 
Extension of holographic dictionary to massive scalar fields is left fo a future work.

\end{document}